\documentstyle[12pt]{article}
\date{}
\newtheorem{Proposition}{Proposition}
\newtheorem{Theorem}{Theorem}
\newtheorem{Lemma}{Lemma}
\newcommand\al{\alpha}

\newcommand\eq{equation}

\newcommand\btd{\raise 2pt
\hbox{$\hat\bigtriangledown$}\hskip 1.5pt}
\newcommand\bt{\raise 2pt
\hbox{$\bigtriangledown$}\hskip 1.5pt}
\begin{document}
\topskip=15mm \baselineskip=15 pt

\begin{center}{ \large\bf The q-deformation of AKNS-D
Hierarchy}
\end{center}

\vspace{0.2in}
\begin{center}
Wang Shikun${}^{1}$\quad Wu Ke${}^2$\quad Wu
Xiaoning${}^1$\quad
Yu Delong${}^{1}$
\end{center}
\begin{center}

${}^1$ Institute of Applied Mathematics, P.O.Box
2734,\\
\quad Academia Sinica, Beijing, China, 100080\\
${}^2$ Institute of Theoretical Physics, P.O.Box
2735,\\
\quad Academia Sinica, Beijing, China, 100080
\end{center}
\par\
\begin{abstract}
In this paper, we considered the q-deformation of
AKNS-D
hierarchy, proved the bilinear identity and give out
the
$\tau$-function of the q-deformed AKNS-D hierarchy.
\end {abstract}

\section{Introduction}
Recently more and more attentions have been paid to
the discrete
integrable system. For example one can find nice
review and lot of
references about discrete integrable system in the
book \cite{bob}
edited by Bobenko and Seiler. One of which is the
semi-discretization of integrable system. It is an
integrable
system with many variables and one discrete variable.
There are
lot of discussions about the q-deformation of KdV and
KdV
hierarchy, q-deformation of KP hierarchy, their
solutions and
$\tau$-functions as well as other properties
\cite{zhang}-\cite{tu2}. Where $\tau$-functions of the
q-KP
hierarchy \cite{il1},\cite{il2} could be constructed
from the
$\tau$-functions of classical KP hierarchy by making
an
appropriate shift. In this paper, we consider the
AKNS-D hierarchy
which is proposed by Dickey in Ref.\cite{dk2} and its
q-deformation.

This paper is organized as follow : In section II, we
will give a
brief review of the AKNS-D hierarchy to make this
paper
self-contained. In section III, we will give out the
definition of
the q-AKNS-D hierarchy, their Baker functions and the
bilinear
identity. The derivation of the $\tau$-functions are
placed in
section IV.

\section{Brief Review of the AKNS-D Hierarchy}
In this section, we briefly review the general idea of
AKNS-D
hierarchy. The details can be found in
Ref.\cite{dk1},\cite{dk2},\cite{dk3}.

\subsection{Definition}
Let
\begin{equation}
\label{L} L=\partial +U- zA
\end{equation}
where $A=diag(a_1,a_2,\cdots,a_n)$ , $U$ is an n-dim
matrix
function and $u_{ii}=0$. Resolvent of $L$ is defined
as a series :
\begin{\eq}
\label{R} R=\sum^{\infty}_{i=0}R^{(i)}z^{-i}
\end{\eq}
which is communitive with $L$, i.e. $[L,R]=0$.The
elements of R
are all differential polynomials of $u_{ij}$. All of
the
Resolvents form an n-dim algebra over the field of
constant
diagonal series $C(z)=\sum^{\infty}_{i_1}C_i
z^{-i}$.The basis of
Resolvents are ${R_{\alpha}}$
\begin{\eq}
\label{rb}
R_{\alpha}=\sum^{\infty}_{j=0}R_{\alpha}^{(j)}z^{-j}
\end{\eq}
where $R_{\alpha}^{(0)}=E_{\alpha}$ and $E_{\alpha}$
is the matrix
with the only non-zero element at the $\alpha\!\alpha$
place. All
elements of $R^{(j)}, j>0$ are differential
polynomials without
constants. The basic resolvents satisfy the relation
$R_{\alpha}R_{\beta}=\delta_{\alpha\beta}R_{\beta}$.
Take
\begin{\eq}
\label{B} B_{k\alpha}=(z^k
R_{\alpha})_+=\sum^{k}_{j=0}R_{\alpha}^{(j)}z^{k-j}
\end{\eq}
The subscript + means taking non-negative powers of
$z$. The
hierarchy is the set of equations
\begin{\eq}
\label{akns}
\partial_{k\alpha}L=[B_{k\alpha}, L]
\end{\eq}
where $\partial_{k\alpha}$ means
$\frac{\partial}{\partial
t_{k\alpha}}$, and $t_{k\alpha},\ k=0,1,2,\cdots,\
\alpha=1,2,\cdots,n$ is a set of time variables. In
these
equations, the variables $\partial$ and
$\partial_{k\alpha}$ are
not independent. They have the following relation :
\begin{\eq}
\label{xt}
\partial=\sum_{\alpha=1}^{n}a_{\alpha}
\partial_{1\alpha}
\end{\eq}

\subsection{Dressing method}
Define
\begin{\eq}
\hat w(z)=I+\sum^{\infty}_{j=1} w_j z^{-j}
\end{\eq}
The operator $L$ can be represented as the following
form
\begin{\eq}
L=\hat w(z)(\partial-zA)\hat w(z)^{-1}
\end{\eq}
The basic resolvents is given by
\begin{\eq}
R_{\alpha}=\hat w(z) E_{\alpha}\hat w(z)^{-1}
\end{\eq}
The formal Baker function is
\begin{\eq}
w=\hat
w(z)exp(\sum^{\infty}_{k=0}\sum^{n}_{\alpha=0}z^kE_{\alpha}t_{k\alpha})
\end{\eq}
then we have
\begin{eqnarray}
L&=&w\partial w^{-1}\\
R_{\alpha}&=&wE_{\alpha}w^{-1}
\end{eqnarray}
The equation of the hierarchy are equivalent to
\begin{eqnarray}
L(w)&=&0\nonumber\\
\partial_{k\alpha}w&=&B_{k\al}w\quad or\quad
\partial_{k\al}
\hat w=-(z^kR_{\al})_-\hat w
\end{eqnarray}
To avoid some passible confusions, from now on we use
$L(f)$ or
$(Lf)$to note an operator $L$ act on a function $f$
and use $Lf$
to note two operator's multiplication.

\subsection{Bilinear Identity and $\tau$-function}
The adjoint Baker function is defined as
$w^*=(w^{-1})^T$ which
satisfies the adjoint equation
\begin{\eq}
L^*(w^*)=0
\end{\eq}
where $L^*=-\partial +(U-zA)^T$.

\begin{Proposition}(Bilinear Identity)

A bilinear relation
\begin{displaymath}
res_z[z^l(\partial_{k_1\al_1}\cdots\partial_{k_s\al_s}w)(w^*)^T]=0
\end{displaymath}
holds, where $l=0,1,\cdots$ and
$(k_1\al_1),\cdots,(k_s,\al_s)$ is
any set of indeces.Conversly, let two functions
\begin{displaymath}
w=(I+\sum^{\infty}_{i=1}w_iz^{-i})exp(\sum^{\infty}_{k=0}
\sum^{n}_{\al=1}z^kE_{\al}t_{k\al})
\end{displaymath}
\begin{displaymath}
w^*=(I+\sum^{\infty}_{i=1}w^*_iz^{-i})exp(-\sum^{\infty}_{k=0}
\sum^{n}_{\al=1}z^kE_{\al}t_{k\al})
\end{displaymath}
satisfy above bilinear identity,then $w$ and $w^*$ are
the baker
function and the adjoint baker function of an operator
$L$ which
satisfies the hierarchy equation.
\end{Proposition}

\section{q-AKNS-D hierarchy}

\subsection{some useful results of q-calculation}
In this section, we will give some basic definitions
about
q-difference calculation and some useful relations
without proof.
Details can be found in Ref.\cite{il2},\cite{gr}.

First, we introduce two operators :
\begin{eqnarray}
Df(x)&:=&f(qx)\\
D_qf(x)&:=&\frac{f(qx)-f(x)}{x(q-1)}
\end{eqnarray}
The first difference between q-difference and ordinary
differential calculus is the Leibnitz's law. The
q-Leibnitz's law
is
\begin{eqnarray}
D_q(fg)&=&(Df)\cdot(D_qg)+(D_qf)\cdot g\nonumber\\
{ }&=&f\cdot(D_qg)+(D_qf)\cdot(Dg)
\end{eqnarray}
Using this Leibnitz's law, it is easy to show the
following lemma
\begin{Lemma}
\begin{eqnarray}
D_q^mD_q^nf&=&D_q^{m+n}f
\end{eqnarray}
\end{Lemma}
Another useful fact about q-difference is
q-exponential
function.It is defined as
\begin{equation}
exp_q (x):=\sum^{\infty}_{k=0}\frac{(1-q)^k}{(q;
q)_k}x^k
\end{equation}
where $(a; q)_k:=\prod^{k-1}_{s=0}(1-aq^s)$ and $(a;
q)_0=1$. The
useful feacture of this function is that the behavior
of
$exp_q(x)$ acted by q-difference operator is just like
the
exponential function acted by the ordinary
differential operator,
i.e.
\begin{equation}
D_qexp_q(zx)=zexp_q(zx)
\end{equation}
Other two useful relations about $exp_q(x)$ are :
\begin{eqnarray}
exp_q(x)&=&exp(\sum^{\infty}_{k=1}\frac{(1-q)^k}{k(1-q^k)}x^k)\\
(exp_q(x))^{-1}&=&exp_{1/q}(-x)
\end{eqnarray}
Using above relation, direct calculation gives
\begin{Lemma}
\label{expqo}
\begin{displaymath}
exp_q(zAx)exp(\sum_{k=0}^{\infty}\sum_{\alpha=1}^{n}z^kE_{\alpha}t_{k\alpha})
=exp(\sum_{k=0}^{\infty}\sum_{\alpha=1}^{n}z^kE_{\alpha}{t'}_{k\alpha})
\end{displaymath}
where
${t'}_{k\alpha}=t_{k\alpha}+\frac{(1-q)^k}{k(1-q^k)}(a_{\alpha}x)^k$
\end{Lemma}
For later convenience, we define the q-commutater ${[\
,\ ]}_q$ as
\begin{equation}
{[A, B]}_q=(DA)\cdot B-B\cdot A
\end{equation}
This bracket can be seen as the q-deform of the
ordinary
commutater. The operator D is coming from the
q-Leibnitz's law.

Introduce a $L^2$-metric on the function space as
\begin{equation}
<A , B>:=tr\int^{+\infty}_{-\infty}A\cdot B dx
\end{equation}
Using this inner product, we can define the dual
operator as usual
\begin{equation}
<f, (g)P^*>:=<P(f), g>
\end{equation}
It is easy to show that
$(D_q)^*=(-\frac{1}{q})D_{1/q}$ and we can
prove that
\begin{Lemma}
Let
\begin{displaymath}
P=\sum_i p_i D^i_q \quad and \quad Q=\sum_j g_j D^j_q
\end{displaymath}
be two q-pseudo-difference operator. Define
\begin{displaymath}
Q|_{x/q}=\sum_j g_j(\frac{x}{q})q^j D^j_q
\end{displaymath}
then we have
\begin{displaymath}
res_z(Pexp_q(zAx)\cdot
exp_{1/q}(-zAx)Q^*|_{x/q})=res_{D_q}(PA^{-1}Q)
\end{displaymath}
\end{Lemma}
Proof : Let $Q^*=\sum_i D^j_{1/q} g_j$, we have
\begin{eqnarray}
&&Pexp_q(zAx)exp_{1/q}(-zAx)Q^*|_{x/q}\nonumber\\
&&=\sum_k\sum_l(p_k(x)D^k_q
exp_q(zAx))(exp_{1/q}(-zAx)D_{1/q}^lq^lg_l
(\frac{x}{q})\nonumber\\
&&=\sum_k\sum_lp_k(x)(Az)^kexp_q(zAx)exp_{1/q}(-zAx)(-Az)^lq^lg_l
(\frac{x}{q})\nonumber\\
&&=\sum_k\sum_l(-q)^lp_k(x)A^{k+l}g_l(\frac{x}{q})z^{k+l}\nonumber
\end{eqnarray}
So we get
\begin{eqnarray}
&&res_z(Pexp_q(zAx)exp_{1/q}(-zAx)Q^*|_{x/q})\nonumber\\
&&=\sum_{k+l=-1}q^lp_k(x)A^{-1}g_l(\frac{x}{q})\nonumber
\end{eqnarray}
Direct calculation gives
\begin{eqnarray}
&&res_{D_q}(PA^{-1}Q)\nonumber\\
&&=\sum_{k+l=-1}(-q)^lp_k(x)A^{-1}g_l(\frac{x}{q})\nonumber
\end{eqnarray}
That finishes the proof.$\Box$

\subsection{q-AKNS-D hierarchy}
Let $L_q=D_q-zA+U$, where
$A=diag(a_1,a_2,\cdots,a_n)$, $U$ is an
n-dim matrix function with $u_{ii}=0$, for any $i$.
Like the
AKNS-D hierarchy, we define the resolvent $R$ for
$L_q$ as
\begin{eqnarray}
R&=&\sum^{\infty}_{i=0}R_i z^{-i}\nonumber\\
{[R, L_q]}_q&=&0
\end{eqnarray}
i.e.
\begin{equation}
\label{qr} D_qR-[R, (U-zA)]_q=0
\end{equation}
Take the formal expression of $R$ into equation
(\ref{qr}), we can
get
\begin{eqnarray}
D_qR^{(j)}-[R^{(j)}, U]_q+[R^{(j+1)},
A]_q=0\label{jorder}\\
{[R^{(0)}, A]}_q=0\label{0order}
\end{eqnarray}

\begin{Theorem}
All of the resolvents R form an algebra over the field
of the
formal series $c(z)=\sum^{\infty}_{i=0}c_i z^{-i}$, we
note it as
$\Re$.
\end{Theorem}
Proof : First, it is easy to see that if $R^1$, $R^2$
satisfy the
equation (\ref{qr}), then we have
$[c_1(z)R^1+c_2(z)R^2, L_q]_q=0$

Second, if $R^1$, $R^2$ are two resolvents, then
\begin{displaymath}
{[R^1R^2, L_q]}_q=(DR^1)[R^2, L_q]_q+[R^1, L_q]_qR^2=0
\end{displaymath}
So all of the resolvents form an algebra.$\Box$

We can define
\begin{equation}
\label{what} \hat w_q:=I+\sum^{\infty}_{k=1}w_kz^{-k}
\end{equation}
Which satisfies
\begin{equation}
L_q=(D\hat w_q)\cdot (D_q-zA)\cdot\hat w_q^{-1}
\end{equation}
The existence of $\hat w_q$ is obvious, because we can
rewrite
above equation as
\begin{equation}
L_q\hat w_q=(D\hat w_q)\cdot (D_q-zA)
\end{equation}
Using the form extension (\ref{what}), we can solve
these $\hat
w_j$ at least formally order by order just like what
we do in the
classical case.
\begin{Theorem}
$R_{\alpha}=\hat w_qE_{\alpha}\hat w_q^{-1}$ is a
resolvent and
has the following property
\begin{displaymath}
R_{\alpha}\cdot
R_{\beta}=\delta_{\alpha\beta}R_{\beta}
\end{displaymath}
\end{Theorem}
Proof :
\begin{eqnarray}
&&{[R_{\alpha}, L_q]}_q\nonumber\\
&&=(D\hat w_q)E_{\alpha}(D\hat w_q^{-1})\cdot L_q
-L_q\cdot\hat w_q E_{\alpha}\hat w_q^{-1}\nonumber\\
&&=(D\hat w_q)E_{\alpha}(D\hat w_q^{-1})\cdot(D\hat
w_q)(D_q-zA)
\hat w_q^{-1}\nonumber\\
&& -(D\hat w_q)(D_q-zA)\hat w_q^{-1}\cdot\hat
w_qE_{\alpha}
\hat w_q^{-1}\nonumber\\
&&=0\nonumber
\end{eqnarray}
and
\begin{eqnarray}
R_{\alpha}R_{\beta}&=&\hat w_qE_{\alpha}\hat
w_q^{-1}\cdot
\hat w_qE_{\beta}\hat w_q^{-1}\nonumber\\
{ }&=&\hat w_qE_{\alpha}E_{\beta}\hat
w_q^{-1}\nonumber\\
{ }&=&\delta_{\alpha\beta}R_{\beta}\qquad \qquad
\qquad\Box\nonumber
\end{eqnarray}

Define $B_{k\alpha}=(z^kR_{\alpha})_+$, $\bar
B_{k\alpha}=(z^kR_{\alpha})_-$ then we define the
q-AKNS-D
hierarchy in the Lax pair form
\begin{eqnarray}
\label{lax}
L_q\hat w_q=(D\hat w_q)(D_q-zA)\nonumber\\
\partial_{k\alpha}\hat w_q=-\bar B_{k\alpha}\hat w_q
\end{eqnarray}
\begin{Theorem}
\label{lp} Above definition can give out following
equation
\begin{displaymath}
\partial_{k\alpha}L_q={[B_{k\alpha}, L_q]}_q
\end{displaymath}
\end{Theorem}
Proof : Because $L_q=(D\hat w_q)(D_q-zA)(\hat
w_q^{-1})$, then
\begin{eqnarray}
&&\partial_{k\alpha}L_q=(\partial_{k\alpha}D\hat
w_q)(D_q-zA)
\hat w_q^{-1}\nonumber\\
&&+(D\hat w_q)(D_q-zA)(\partial_{k\alpha}\hat
w_q^{-1})\nonumber\\
&&=-(D\bar B_{k\alpha})\cdot(D\hat w_q)(D_q-zA)(\hat
w_q^{-1})\nonumber\\
&&+(D\hat w_q)(D_q-zA)\hat w_q^{-1}\bar
B_{k\alpha}\nonumber\\
&&={[-\bar B_{k\alpha}, L_q]}_q\nonumber\\
&&={[B_{k\alpha}-z^kR_{\alpha}, L_q]}_q\nonumber\\
&&={[B_{k\alpha}, L_q]}_q\qquad\qquad\Box\nonumber
\end{eqnarray}
\begin{Theorem}
\begin{displaymath}
\partial_{k\alpha}R_{\beta}=[B_{k\alpha}, R_{\beta}]
\end{displaymath}
\end{Theorem}
Proof :
\begin{eqnarray}
&&\partial_{k\alpha}R_{\beta}=\partial_{k\alpha}(\hat
w_qE_{\beta})
\hat w_q^{-1}\nonumber\\
&&=-\bar B_{k\alpha}\hat w_qE_{\beta}\hat w_q^{-1}
+\hat w_qE_{\beta}\hat w_q^{-1}\bar
B_{k\alpha}\nonumber\\
&&=-[\bar B_{k\alpha}, R_{\beta}]\nonumber\\
&&=[B_{k\alpha}, R_{\beta}]\qquad\qquad\Box\nonumber
\end{eqnarray}
Using this relation, we can easily prove that
$\partial_{k\alpha}B_{l\beta}
-\partial_{l\beta}B_{k\alpha}=[B_{k\alpha},
B_{l\beta}]$.

\begin{Lemma}
Every $R$ can be fixed by its zero order term and
$R_{\alpha}$
form a basis of $\Re$.
\end{Lemma}
Proof : Using Eq.(\ref{0order}), we can get $R^{(0)}$
must be a
diagonal matrix. (Here we require all functions that
we deal with
are bounded and continuous everywhere as the function
of x)
Because of Eq.(\ref{jorder}), we can solve every
$R^{(j)}$ order
by order, the only freedom left is the constant
diagonal part of
$R^{(j)}$ which can be choose as zero. So the linear
independent
solutions are those whose zero order term is
$E_{\alpha}$, that is
why $R_{\alpha}$ form a basis of $\Re$. $\Box$
\begin{Lemma}
We can equivalently define the q-AKNS-D hierarchy as
\begin{displaymath}
\partial_{k\alpha}L_q={[B_{k\alpha}, L_q]}_q
\end{displaymath}
\end{Lemma}
Proof :The Theorem \ref{lp} show that the Lax pair
definition can
lead to above equation, what we need to do is to prove
the
converse direction. For $\forall\ \alpha,\ \beta$,
\begin{eqnarray}
&&{[\partial_{k\alpha}R_{\beta}-[B_{k\alpha},
R_{\beta}], L_{q}]}_q\nonumber\\
&&={[\partial_{k\alpha}R_{\beta}, L_q]}_q
-{[[B_{k\alpha},
R_{\beta}], L_{q}]}_q\nonumber
\end{eqnarray}
But
\begin{eqnarray}
&&{[\partial_{k\alpha}R_{\beta}, L_q]}_q\nonumber\\
&&=\partial_{k\alpha}{[R_{\beta}, L_q]}_q-{[R_{\beta},
\partial_{k\alpha}L_q]}_q\nonumber\\
&&=-{[R_{\beta}, \partial_{k\alpha}L_q]}_q\nonumber\\
&&=-{[R_{\beta}, {[B_{k\alpha}, L_q]}_q]}_q\nonumber\\
&&={[[R_{\beta}, B_{k\alpha}], L_q]}_q\nonumber
\end{eqnarray}
which gives
\begin{displaymath}
{[\partial_{k\alpha}R_{\beta}-[B_{k\alpha},
R_{\beta}],
L_{q}]}_q=0
\end{displaymath}
that means $\partial_{k\alpha}R_{\beta}-[B_{k\alpha},
R_{\beta}]$
is a resolvent. We have known that $R_{\alpha}$ form a
basis of
$\Re$, then we can express
$\partial_{k\alpha}R_{\beta}-[B_{k\alpha}, R_{\beta}]$
as
\begin{displaymath}
\partial_{k\alpha}R_{\beta}-[B_{k\alpha},
R_{\beta}]=\sum_{\alpha}c_{\alpha}(z)
R_{\alpha}
\end{displaymath}
Take $\gamma\ne\beta$, we have
\begin{eqnarray}
&&(\partial_{k\alpha}R_{\beta}-[B_{k\alpha},
R_{\beta}])R_{\gamma}\nonumber\\
&&=-R_{\beta}\partial_{k\alpha}R_{\gamma}+R_{\beta}B_{k\alpha}
R_{\gamma}\nonumber\\
&&=-R_{\beta}\partial_{k\alpha}R_{\gamma}+R_{\beta}B_{k\alpha}R_{\gamma}
-R_{\beta}R_{\gamma}B_{k\alpha}\nonumber\\
&&=-R_{\beta}(\partial_{k\alpha}R_{\gamma}-[B_{k\alpha},
R_{\gamma}])\nonumber\\
&&=-\tilde c_{\beta}R_{\beta}\nonumber
\end{eqnarray}
On the other hand
\begin{displaymath}
(\partial_{k\alpha}R_{\beta}-[B_{k\alpha},
R_{\beta}])R_{\gamma}=\sum_{\alpha}
c_{\alpha}R_{\alpha}R_{\gamma}=c_{\gamma}R_{\gamma}
\end{displaymath}
Because $R_{\gamma}$, $R_{\beta}$ are linear
independent, we get
$c_{\beta} =\tilde c_{\gamma}=0$, so
$\partial_{k\alpha}R_{\beta}-[B_{k\alpha}, R_{\beta}]
=c(z)R_{\beta}$. But
\begin{eqnarray}
&&(\partial_{k\alpha}R_{\beta}-[B_{k\alpha},
R_{\beta}])R_{\beta}\nonumber\\
&&=\partial_{k\alpha}R_{\beta}^2-R_{\beta}\partial_{k\alpha}R_{\beta}
-B_{k\alpha}R_{\beta}+R_{\beta}B_{k\alpha}R_{\beta}\nonumber\\
&&=\partial_{k\alpha}R_{\beta}-B_{k\alpha}R_{\beta}+R_{\beta}B_{k\alpha}
-R_{\beta}\partial_{k\alpha}R_{\beta}-R_{\beta}^2B_{k\alpha}
+R_{\beta}B_{k\alpha}R_{\beta}\nonumber\\
&&=\partial_{k\alpha}R_{\beta}-[B_{k\alpha},
R_{\beta}]
-R_{\beta}(\partial_{k\alpha}R_{\beta}-[B_{k\alpha},
R_{\beta}])\nonumber\\
&&=0\nonumber
\end{eqnarray}
So we get
\begin{displaymath}
\partial_{k\alpha}R_{\beta}-[B_{k\alpha}, R_{\beta}]=0
\end{displaymath}
Using this result, it is easy to show
\begin{displaymath}
\partial_{k\alpha}B_{l\beta}-\partial_{l\beta}B_{k\alpha}
=[B_{k\alpha}, B_{l\beta}]
\end{displaymath}
and then we can just follow the standard way to extend
the
operator $\partial_{k\alpha}$ on $w_q$ by requiring
$\partial_{k\alpha}w_q=B_{k\alpha}w_q$ and above
result guarantees
$[\partial_{k\alpha}, \partial_{l\beta}]=0$ hold, so
we prove the
converse direction.$\Box$

\subsection{Baker function and bilinear identity of
the q-AKNS-D hierarchy}
Define the Baker function as
\begin{equation}
\label{baker} w_q=\hat w_q\cdot
exp_q(zAx)exp(\sum_{k=1}^{\infty}\sum_{\alpha=1}^{n}
z^kE_{\alpha}t_{k\alpha})
\end{equation}
\begin{Theorem}
\begin{eqnarray}
\label{w}
L_q(w_q)&=&0\nonumber\\
\partial_{k\alpha}w_q&=&B_{k\alpha}w_q
\end{eqnarray}
\end{Theorem}
Proof : Using the Eq.(\ref{lp}), we get
\begin{eqnarray}
&&L_q(w_q)=L_q(\hat w_q\cdot
exp_q(zAx)exp(\sum_{k=1}^{\infty}
\sum_{\alpha=1}^{n}
z^kE_{\alpha}t_{k\alpha}))\nonumber\\
&&=D\hat
w_q(D_q-zA)exp_q(zAx)exp(\sum_{k=1}^{\infty}\sum_{\alpha=1}^{n}
z^kE_{\alpha}t_{k\alpha})\nonumber\\
&&=0\nonumber
\end{eqnarray}
\begin{eqnarray}
&&\partial_{k\alpha}w_q=\partial_{k\alpha}(\hat
w_q\cdot
exp_q(zAx) exp(\sum_{k=1}^{\infty}\sum_{\alpha=1}^{n}
z^kE_{\alpha}t_{k\alpha}))\nonumber\\
&&=-\bar B_{k\alpha}\hat w_q\cdot
exp_q(zAx)exp(\sum_{k=1}^{\infty} \sum_{\alpha=1}^{n}
z^kE_{\alpha}t_{k\alpha})\nonumber\\
&&+\hat
w_qz^kE_{\alpha}exp_q(zAx)exp(\sum_{k=1}^{\infty}\sum_{\alpha=1}^{n}
z^kE_{\alpha}t_{k\alpha})\nonumber\\
&&=-\bar B_{k\alpha}\cdot w_q+z^kR_{\alpha}\cdot
w_q\nonumber\\
&&=B_{k\alpha}\cdot w_q\nonumber
\end{eqnarray}
\begin{Theorem}
(Bilinear Identity)

 i) If $w_q$ is a solution of Eq.(\ref{w}), it
satisfies the following
 identity
 \begin{equation}
 \label{qb1}
res_z(z^l(D_q^m\partial_{k\alpha}^{[\lambda]}w_q)\cdot
w_q^{-1})=0
 \end{equation}
 for $l=0,1,2,\cdots$, $m=0,1$,
 $\forall [\lambda]$(where

$\partial_{k\alpha}^{[\lambda]}=\partial_{k_1\alpha_1}\partial_{k_2\alpha_2}
 \cdots\partial_{k_s\alpha_s}$, and
$(k_1,\alpha_1),\cdots,(k_s,\alpha_s)$ is
 any set of indices).

 ii) If
 \begin{eqnarray}
w_q&=&(I+\sum_{i=1}^{\infty}w_iz^{-i})exp_q(zAx)exp(\sum_{k=1}^{\infty}\sum_{\alpha=1}^{n}
z^kE_{\alpha}t_{k\alpha})\nonumber\\
w_q^*&=&(I+\sum_{i=1}^{\infty}w_i^*z^{-i})exp_{1/q}(-zAx)
exp(-\sum_{k=1}^{\infty}\sum_{\alpha=1}^{n}z^kE_{\alpha}t_{k\alpha})\nonumber
\end{eqnarray}
and they satisfy the following identity
\begin{equation}
\label{qb2}
res_z(z^l{D_q^m\partial_{k\alpha}^{[\lambda]}w_q}\cdot
(w_q^*)^t)=0
\end{equation}
for $l=0,1,2,\cdots,\ m=0,1$, $\forall [\lambda]$,
then we have

1) $(w_q^{-1})^t=w_q^*$.

2) $w_q$ is a solution of Eq.(\ref{w})
\end{Theorem}
Proof :

i) If $w_q$ is a solution of the q-AKNS hierarchy, we
have
\begin{displaymath}
\partial_{k\alpha}w_q=B_{k\alpha}w_q
\end{displaymath}
So
$\partial_{k\alpha}^{[\lambda]}w_q=f(B_{k\alpha})\cdot
w_q$,
where $f(B_{k\alpha})$ is a differential polynomial of
$B_{k\alpha}$ and it is easy to see that
$(f(B_{k\alpha}))_+=f(B_{k\alpha})$. Furthermore, we
also know
$D_qw_q=(zA-U)\cdot w_q$, then for $\forall l\ge 0$
and $\forall
[\lambda]$
\begin{eqnarray}
&&res_z(z^l{D_q\partial_{k\alpha}^{[\lambda]}w_q}\cdot
(w_q)^{-1})\nonumber\\
&&=res_z(z^l(D_qf(B_{k\alpha}))w_q\cdot w_q^{-1})
+res_z(z^l(Df(B_{k\alpha}))\cdot (zA-U)w_q\cdot
w_q^{-1})\nonumber\\
&&=res_z(z^l(D_qf(B_{k\alpha}))+z^l(Df(B_{k\alpha}))(zA-U))\nonumber\\
&&=0\nonumber
\end{eqnarray}
ii) First, choosing $m=[\lambda]=0$, Eq.(\ref{qb2})
gives out
\begin{displaymath}
res_z(z^lw_q\cdot (w_q^*)^t)=0
\end{displaymath}
for $\forall l\ge 0$. This means that $w_q\cdot
(w_q^*)^t$ doesn't
contain negative power term. From the formal expresion
of $w_q$
and $w_q^*$, we know it also doesn't contain positive
power term
and the zero order term is $I$, so we get
$(w_q^{-1})^t=w_q^*$.

Second,
\begin{eqnarray}
&&\partial_{k\alpha}w_q-B_{k\alpha}w_q\nonumber\\
&&=(\partial_{k\alpha}\hat
w_q)exp_q(zAx)exp(\sum_{k=1}^{\infty}\sum_{\alpha=1}
^{n}z^kE_{\alpha}t_{k\alpha})\nonumber\\
&&+\hat w_qexp_q(zAx)
exp(\sum_{k=1}^{\infty}\sum_{\alpha=1}^{n}z^kE_{\alpha}t_{k\alpha})z^kE_{\alpha}
-(z^kR_{\alpha})_+w_q\nonumber\\
&&=(\partial_{k\alpha}\hat
w_q)exp_q(zAx)exp(\sum_{k=1}^{\infty}\sum_{\alpha=1}
^{n}z^kE_{\alpha}t_{k\alpha})+z^kw_qE_{\alpha}w_q^{-1}\cdot
w_q
-(z^kR_{\alpha})_+w_q\nonumber\\
&&=(\partial_{k\alpha}\hat w_q+(z^kR_{\alpha})_-\hat
w_q)exp_q(zAx)
exp(\sum_{k=1}^{\infty}\sum_{\alpha=1}^{n}z^kE_{\alpha}t_{k\alpha})\nonumber
\end{eqnarray}
From Eq.(\ref{qb2}), we have
\begin{displaymath}
res_z(z^l(\partial_{k\alpha}-B_{k\alpha})w_q\cdot
(w_q^*)^t)=0
\end{displaymath}
But
\begin{displaymath}
(\partial_{k\alpha}-B_{k\alpha})w_q\cdot w_q^{-1}=
(\partial_{k\alpha}\hat w_q+(z^kR_{\alpha})_-\hat
w_q)\cdot \hat
w_q^{-1}
\end{displaymath}
So we can see it contains no positive power term and
the bilinear
identity is
\begin{displaymath}
res_z(z^l(\partial_{k\alpha}\hat
w_q+(z^kR_{\alpha})_-\hat
w_q)\cdot \hat w_q^{-1})=0
\end{displaymath}
for $\forall l\ge 0$, which gives out
\begin{displaymath}
\partial_{k\alpha}\hat w_q+(z^kR_{\alpha})_-\hat w_q=0
\end{displaymath}
and we have known it is equavilent to equation
\begin{displaymath}
\partial_{k\alpha}w_q=B_{k\alpha}w_q
\end{displaymath}
Define : $L_q=(Dw_q)D_qw_q^{-1}$. Simple calculation
gives
\begin{displaymath}
L_q=D_q-(D_qw_q)\cdot w_q^{-1}
\end{displaymath}
Using the formal expression of $w_q$, direct
calculation gives the
highest order of $D_qw_q\cdot w_q^{-1}$ is $zA$ and
the bilinear
identity gives that $D_qw_q \cdot w_q^{-1}$ doesn't
contain
negative power term, we note the zero order term of
$D_qw_q\cdot
w_q^{-1}$ as $-U$, then we get
\begin{displaymath}
L_q=D_q+zA-U
\end{displaymath}
So such $w_q$ satisfies Eq.(\ref{w}). $\Box$

\section{The $\tau$-function of q-AKNS-D hierarchy}
In Ref.\cite{il2}, P.Iliev gives a way to construct
the
$\tau$-function of the q-KP hierarchy. The main idea
is to
"q-shift" the time variables $t_{k\alpha}$ of the
classical KP
hierarchy's $\tau$-function and prove that the Baker
function
construct from this kind of $\tau$-function satisfies
the bilinear
identity of the q-KP hierarchy. In this section, we
will this way
and generalize P.Iliev's method to the q-AKNS-D
hierarchy.

Define : the q-shift of $t_{k\alpha}$ is defined as
\begin{displaymath}
t_{k\alpha} \longmapsto
t_{k\alpha}+\frac{(1-q)^k}{k(1-q^k)}(a_{\alpha}x)^k
\end{displaymath}
and for convenience, we also note as $t+{[Ax]}_q$.
\begin{Theorem}(The $\tau$-function of q-AKNS-D
hierarchy)

If $\tau (t)$ is a $\tau$-function of the classical
AKNS-D
hierarchy, then
\begin{displaymath}
\tau_q(t;\ x):=\tau (t+{[Ax]}_q)
\end{displaymath}
is a $\tau$-function of q-AKNS-D hierarchy.
\end{Theorem}
Proof : In Ref.\cite{dk2}, Dickey gives out some
$\tau$-functions
of AKNS-D hierarchy. To construct the Baker function,
they have
the following form :
\begin{eqnarray}
\label{swtw} &&\hat
w_{W\alpha\beta}(t,z)=z^{-1}\frac{\tau_{W\alpha\beta}
(\cdots,t_{k\beta}-\frac{1}{k}z^{-k},\cdots)}
{\tau_W(t)}\nonumber\\
&&\hat w_{W\alpha\alpha}(t,z)=\frac{\tau_W
(\cdots,t_{k\alpha}-\frac{1}{k}z^{-k},\cdots)}
{\tau_W(t)}
\end{eqnarray}
which is simuler to the KP hierarchy. (more detials
can be found
in Ref.\cite{dk2},\cite{dk3},\cite{sw}.)

Because we have defined the q-$\tau$-function as
\begin{displaymath}
\tau_q(t;\ x):=\tau(t+{[Ax]}_q)
\end{displaymath}
Using Eq.(\ref{swtw}), (\ref{baker}) and Lemma
\ref{expqo}, it is
easy to see that the q-Baker function $w_q$ is just
the q-shift of
the classical Baker function, i.e.
\begin{displaymath}
w_q(t;\ x)=w(t+{[Ax]}_q)
\end{displaymath}
the classical Baker function $w$ satisfies the
classical bilinear
identity
\begin{equation}
\label{cb}
res_z(z^l(\partial_{k\alpha}^{[\lambda]}w)\cdot
w^{-1})=0 \qquad for\ \forall\ l\ge 0,\ \forall\
[\lambda]
\end{equation}
we want to show that the function $w_q$ satisfies the
q-bilinear
identity
\begin{displaymath}
res_z(z^l(D_q\partial_{k\alpha}^{[\lambda]}w_q)\cdot
w_q^{-1})=0
\end{displaymath}
Take the expression of $w_q$ into the q-bilinear
identity, we get
\begin{eqnarray}
\label{2term}
&&res_z(z^l(D_q\partial_{k\alpha}^{[\lambda]}w(t+{[Ax]}_q))
\cdot w^{-1}(t+{[Ax]}_q)\nonumber\\
&&=[res_z(z^l(\partial_{k\alpha}^{[\lambda]}w(t+{[Aqx]}_q))
\cdot w^{-1}(t+{[Ax]}_q)\nonumber\\
&&-res_z(z^l(\partial_{k\alpha}^{[\lambda]}w(t+{[Ax]}_q))
\cdot
w^{-1}(t+{[Ax]}_q)]\cdot\frac{1}{x(q-1)}
\end{eqnarray}
Choosing $t'=t+{[Ax]}_q$, the classical bilinear
identity
Eq.(\ref{cb}) gives the second term of right side of
above
equation is zero.

The first term of above equation is
\begin{displaymath}
\frac{1}{x(q-1)}[res_z(z^l(\partial_{k\alpha}^{[\lambda]}w(t+{[Aqx]}_q))
\cdot w^{-1}(t+{[Ax]}_q)]
\end{displaymath}
Note  $\frac{(1-q)^k}{k(1-q^k)}(a_{\alpha}qx)^k$  as
$x_{k\alpha}^q$  and
$\frac{(1-q)^k}{k(1-q^k)}(a_{\alpha}x)^k$  as
$x_{k\alpha}$,  take the Taylar extension of
$w(t+{[Ax]}_q)$ at
$t+{[Ax]}_q$, we get
\begin{eqnarray}
\label{talar}
&&\frac{1}{x(q-1)}[res_z(z^l(\partial_{k\alpha}^{[\lambda]}w(t+{[Aqx]}_q))
\cdot w^{-1}(t+{[Ax]}_q))]\nonumber\\
&&=\frac{1}{x(q-1)}[res_z(z^l(\partial_{k\alpha}^{[\lambda]}w(t+{[Ax]}_q))
\cdot w^{-1}(t+{[Ax]}_q))\nonumber\\
&&+\sum_{l,\beta,[\eta]}res_z(z^l(\partial_{k\alpha}^{[\lambda]}
\partial_{l\beta}^{[\eta]}w(t+{[Ax]}_q))
\cdot
w^{-1}(t+{[Ax]}_q))\cdot(x_{l\beta}^q-x_{l\beta})^{[\eta]}]\nonumber\\
\end{eqnarray}
every term in above equation has the following form :
\begin{displaymath}
\frac{(x_{l\beta}^q-x_{l\beta})^{[\eta]}}{x(q-1)}
res_z(z^l(\partial_{k\alpha}^{[\lambda]}\partial_{l\beta}^{[\eta]}w(t+{[Ax]}_q))
\cdot w^{-1}(t+{[Ax]}_q))
\end{displaymath}
where the $\partial_{l\beta}^{[\eta]}$ and the
$(x_{l\beta}^q-x_{l\beta})^{[\eta]}$ are come from the
Taylar
extension. The classical bilinear identity of AKNS
hierarchy make
sure that this kind of term is zero, for any
$[\lambda]$, $[\eta]$
and $l \ge 0$, so we get the first term of
Eq.(\ref{2term}) is
also zero. That means the Baker function $w_q$
satisfies the
q-bilinear identity. $\Box$

Furthermore, comparing both sides of the Taylar
extension of
Eq.(\ref{2term}), we can see that the $D_q$ and the
$\partial_{k\alpha}$ are not independent, we have the
following
relation :
\begin{eqnarray}
D_q&=&\frac{1}{x(q-1)}\sum_{l,\beta,\eta}c([\eta])
(x_{l\beta}^q-x_{l\beta})^{[\eta]}\partial_{l\beta}^{[\eta]}\nonumber\\
{ }&=&\sum_{\beta}a_{\beta}\partial_{1\beta}+O(q-1)
\end{eqnarray}
where $c([\eta])$ is the constant coming from the
Taylar
extension. When $q\to 1$, we can get the relation :
\begin{displaymath}
\partial=\sum_{\beta}a_{\beta}\partial_{1\beta}
\end{displaymath}
which is holed for the classical AKNS hierarchy.

\section*{Acknowledgements}


\begin{thebibliography}{99}

\bibitem{bob} {\em Discrete Integrable Geometry and
Physics}, eds. A.I.Bobenko and R.Seiler,
Oxford University Press, 1999

\bibitem{zhang} D.H.Zhang, J. Phys. A : Math. Gen.
{\bf 26} (1993) 2389

\bibitem{wu} Z.Y.Wu, D.H.Zhang and Q.R.Zheng, J. Phys.
A : Math. Gen. {\bf 27} (1994) 5307

\bibitem{mac} J.Mac and M.Seco, J. Math. Phys. {\bf
37} (1996) 6510

\bibitem{fr} E.Frenkel, Int. Math. Res. Notice {\bf 2}
(1996) 55

\bibitem{fr2} E.Frenkel and N.Reshetkhin, Commun.
Math. Phys. {\bf 178} (1996) 237

\bibitem{kh} B.Khesin, V.Lyubashenko and C.Roger, J.
Func. Anal. {\bf 143} (1997) 55

\bibitem{adler} M.Adler, E.Horozov and P.van Moerbeke,
Phys. Lett. {\bf A 242} (1998) 139

\bibitem{il1} L.Haine and P.Iliev , J. Phys. A : Math.
Gen. {\bf 30} (1997) 7217

\bibitem{il2} P.Iliev , Lett. Math. Phys. {\bf 44}
No.3 (1998) 187

\bibitem{tu1} M.H.Tu, J.C.Shaw and C.R.Lee, Lett.
Math. Phys. {\bf 49} (1999) 33

\bibitem{tu2} M.H.Tu, Lett. Math. Phys. {\bf 49}
(1999) 95

\bibitem{dk1} L.A.Dickey, {\em Soliton Equations and
Hamiltonian System},
              Advanced Series in Mathematical Physics,
Vol.{\bf 12},
              World scientific Press, 1991

\bibitem{dk2} L.A.Dickey, J. Math. Phys. {\bf 32}
(1991) 2996

\bibitem{dk3} L.A.Dickey, On Segal-Wilson's Definition
of the $\tau$-function
              and Hierarchy AKNS-D and mcKP in {\em
Integrable Systems :
              The Verdier Memerial Conference} Ed. by
Olivier Babelon, Pierre
              Cartier and Yvetten Kosmann-Schwarzbach,
              Birkhauser Boston, 1993

\bibitem{sw}  G.Segal and G.Wilson, Publ. Math.
I.H.E.S. {\bf 63} (1985) 1

\bibitem{gr}  G.Gasper and M.Rahman, {\em Basic
Hypergeometric Series},
              Encyclopedia Math. Appl. {\bf 35},
Cambridge University Press,
              1990
\end{thebibliography}
\end{document}